# Input to the European Strategy for Particle Physics

The U.S. Higgs Factory Coordination Consortium
contacts: S. Rajagopalan, T. Raubenheimer

March 31, 2025

## Introduction

This white paper responds to the request by the European Strategy Group (ESG) to submit national inputs as part of the European Strategy for Particle Physics Update (ESPPU). It focuses on CERN's future collider options and provides strong support for FCC-ee as its preferred next major flagship project. The paper follows the ECFA guidelines[1], responding explicitly to item 3, and is supported by recent developments following the 2023 P5 Report[2]. It also outlines the U.S. scientific and engineering expertise including potential technical contributions to a future Higgs Factory.

## Perspective on U.S. - CERN Partnership

With over 2000 members, the United States forms the largest national user community at CERN and is making financial, intellectual, and leadership contributions to a number of projects including LHC/HL-LHC. The U.S.-CERN collaboration[3] and partnership is historical, bound by our common scientific vision and interests, and governed by an international cooperation agreement between the U.S. government agencies and CERN. This collaboration is reciprocal, with CERN and the Organization's member states actively engaged in projects hosted by the United States. As CERN and its member states plan to define the next major flagship project, the historical precedence promises the continuation of the decades-long and trusted U.S.-CERN partnership. The U.S. community is excited to develop the next generation of accelerator and experimental detector technologies and address the compelling science of a Higgs Factory as captured in the 2022 Snowmass Community meeting and the 2023 P5 report. The U.S. eagerly awaits a timely decision by the CERN Council that will propel and solidify the next phase of our partnership.

## Context for this paper

The 2023 report[2] of the Particle Physics Project Prioritization Panel (P5), developed under the purview of the U.S. High Energy Physics Advisory Panel (HEPAP), laid out a compelling scientific argument for an off-shore $e^+e^-$ Higgs Factory and recommended world-leading facilities with new capabilities, as well as maintaining a robust scientific research program.

Following the recommendation of P5, the U.S. Department of Energy (DOE) and the National Science Foundation (NSF) set in place a nationally coordinated U.S. Higgs Factory Coordination Consortium (HFCC)[4] to provide strategic direction and leadership for the U.S. community to engage, shape, and thereby advance the development of the physics, experiment, and detector (PED) and the accelerator (ACC) program for a potential future Higgs factory; and to ensure cooperation with our partners in the international program. Furthermore, DOE and NSF charged the HFCC chairs to lead an Editorial Board[5] and provide input to the ongoing European Strategy Update on a future Higgs Factory located at CERN in consultation with the U.S. community. A series of planning and community meetings[6] were held to seek input from the broader U.S. community, including an American Physical Society (APS) driven DPF/DPB meeting[7]. Additional inputs and endorsements were received from the Laboratory Coordination Groups (LCGs)[8] and the coordinators of the respective PED and ACC technical subsystems within the U.S. HFCC organization. Feedback and endorsement[9] from the U.S. community was sought on the final draft of the paper. This paper summarizes the overall efforts from this process.

**U.S. Vision**

The U.S. has been engaged in developing the physics case and design of a future $e^+e^-$ Higgs Factories for over three decades. The recently concluded P5 process put forward a strong recommendation to encourage the U.S. community to actively engage in the ongoing feasibility studies and subsequently collaborate with international partners to realize an off-shore $e^+e^-$ Higgs Factory. The P5 recommendation also emphasized the timely realization of the off-shore Higgs Factory. Furthermore, P5 suggested a contribution to such a future collider at funding levels commensurate with that of the U.S. involvement in the LHC and HL-LHC, but one that will be subject to future internal U.S. processes and approvals. In 2020, U.S. and CERN signed a bilateral agreement[10] enabling U.S. participation in the Feasibility Study for a future collider. Building on the 2020 agreement, a Statement of Intent (SOI)[11] was signed by the U.S. government and CERN in 2024 which, among other provisions, re-emphasizes the importance of enhancing the collaboration in future planning activities, stating: *"Should the CERN Member States determine the FCC-ee is likely to be CERN's next world-leading research facility following the high-luminosity Large Hadron Collider, the United States intends to collaborate on its construction and physics exploitation, subject to appropriate domestic approvals"*.

The current baseline option, recommended for a feasibility study by the 2020 ESPPU strategy[12] and approved by the CERN Council, is to build FCC-ee followed by FCC-hh (the Integrated FCC program) upon completion of the HL-LHC program. The scientific merits of the Integrated FCC program[13] will address fundamental questions of paramount scientific importance and will ultimately provide crucial tests of the Standard Model and opportunities for new discoveries. The FCC-ee collider is a 90 km synchrotron that is designed to provide high luminosity to four Interaction Points (IP) with an operational energy that ranges from the *Z*-pole to $t\bar{t}$ (90 to 370 GeV). The focus of the design has been to deliver high-luminosity data and collect >$5\times10^{12}$ *Z*'s, >$2\times10^8$ *WW*, >$2\times10^6$ *ZH* and >$2\times10^6$ $t\bar{t}$ events to enable high-precision measurements of SM processes, Higgs boson couplings, and to search for new phenomena and rare processes. A possible additional dedicated operation at the Higgs pole allows probing of the Higgs-Yukawa coupling to electrons via the *s*-channel resonant production. The tunnel would allow the possible future installation of an FCC-hh aiming at the 100 TeV (10 TeV parton Center-of-Mass, or pCM) energy scale. It complements the FCC-ee program by enabling direct discoveries of new particles at unprecedented energy scales and achieves (sub)percent precision measurements of Higgs self-couplings and rare decay modes. Moreover, the FCC-ee design is well established while additional R&D on high-field magnets will be needed to realize a cost-effective FCC-hh program. CERN Council initiated a Feasibility Study[13] of the FCC to evaluate the cost, technical design, and site feasibility; that concludes in March 2025.

In parallel, the Linear Collider (LC) Vision program[14] has begun adapting the ILC design to the CERN site. Such a collider offers $e^+e^-$ collisions up to an energy of 550 GeV with highly polarized beams and can study the *Z*-pole, search for anomalous processes, and offers precision measurements of the top quark, Higgs couplings including the top-Yukawa coupling, as well as the direct measurement of the Higgs self-coupling. The ILC is a well-established collider design based on superconducting RF (SRF) technology which has been demonstrated at the operating 17.5 GeV EuXFEL facility at DESY Hamburg. The LC Visions adaptation further proposes to extend the ILC design to increase its baseline luminosity by a factor of 4; this is a recent modification that needs technical review. The ILC collision energy may be extendable to the TeV-scale within the same tunnel using advanced RF technologies that are currently being developed and possibly to even higher energies using advanced acceleration concepts which are at a pre-conceptual R&D stage.

The options described above are aligned with the vision expressed by P5: "*In 20 years, the HL-LHC program will be completed, a Higgs factory will be preparing to take data, and a vigorous R&D program will be paving the path to a 10 TeV pCM collider. Each of these projects will fill in the map of the Higgs boson's behavior in complementary ways: The HL-LHC will deliver the first draft, the Higgs factory will provide incredible detail in key areas of the landscape, and the 10 TeV pCM collider will reveal the challenging heights of the Higgs boson's interaction with itself.*"

Given the above statements, it is essential to note the following points:
- Strong and enthusiastic interest exists in the U.S. community to advance, together with international partners, an off-shore future Higgs Factory.
- The United States and CERN have a historically strong partnership that continues to deliver major scientific discoveries and technological advances. The U.S. expects to continue its scientific collaboration with CERN and engage with CERN on future approved accelerator and experimental projects.
- P5 has put the completion of the HL-LHC as one of its highest priorities and the physics realized at the HL-LHC could well shape the future direction of particle physics. The U.S. community agrees and strongly supports continuing to give high priority to the successful completion of the HL-LHC program with sufficient resources to collect the planned 3000 fb$^{-1}$ and achieve its physics goals.
- P5 recommended an off-shore Higgs Factory as the next major flagship project for the collider program, adding that both FCC-ee and ILC would meet the scientific requirements. FCC-ee is the current baseline option justified by the 2020 ESPPU strategy, and the U.S. is contributing to the relevant feasibility studies. If FCC-ee is approved by the CERN Council, the U.S. envisions a substantial participation in the FCC-ee experiments and accelerator. If the CERN Council instead chooses construction of a linear collider, which also satisfies P5's scientific goals in a timely manner, the U.S. is expected to similarly participate. In either case, the level and nature of U.S. participation will be determined once a specific project is deemed feasible and well-defined.
- A viable financial model for the next project at CERN is essential and the U.S. participation in any CERN project must take into account the internal U.S. priorities and the need to maintain a healthy U.S. on-shore program in particle physics, one with international participation. This program was enunciated by P5 and includes R&D toward a Muon Collider and aspirations to host an energy frontier collider in the future.
- A clear and prioritized recommendation by the ESG and a prompt subsequent decision by the CERN Council would firmly lay the groundwork for the future of the field, unify the community to work towards a common purpose, and position the field for the decades to come. The need for a timely decision and the rapid realization of physics has been strongly emphasized by the U.S. community, in particular by the early career researchers.

**Strategic discussion on future directions** responding to ECFA Guidelines (item 3)

a) *Which is the preferred next major/flagship collider project for CERN?*

The FCC Feasibility Study has been completed and is now under consideration by the ESG. To date, the FCC feasibility studies have revealed no technical show-stoppers for FCC-ee. Given the comprehensive breadth of the physics program; the technical, cost, and siting maturity of the FCC-ee program; and the need for a timely decision, **we strongly support FCC-ee as the next major flagship project at CERN.**

b) *What are the most important elements in the response to a)?*

A number of elements factor into the decision for the next preferred project. The two most important elements are:

- ***Physics*** must be the driving priority for any current or future programs. It is the primary reason for constructing and operating research facilities.
- A ***timely*** decision and the subsequent realization of physics from the next project are critical. A timely decision is necessary to motivate and guide the next generation of physicists and engineers, maintain the expertise, unify the community, and create a new talent pool. It is also critical to achieve a constant steady flow of scientific results, advance scientific progress, and ensure a seamless and rapid transition from the HL-LHC program to the next flagship project.

Other elements relevant to the decision-making process include sufficient Financial and Human resources to execute the project, the long-term perspective of the physics program, as well as Careers and Training. It has been recognized by several in the U.S. community that a large gap between HL-LHC and the next project would erode the talent pool of engineers and physicists and adversely affect the training of next-generation scientists.

c) *Should CERN/Europe proceed with the preferred option set out in 3a) or should alternative options be considered:*

★ *if Japan proceeds with the ILC in a timely way?*

While an ILC in Japan is consistent with both 2020 ESPPU and the 2023 P5 recommendations, a parallel construction and operation of two large collider projects could become a challenge. ***A reevaluation of the complementarity and the competitiveness of an FCC-ee should be made, if Japan proceeds with the ILC.***

★ *if China proceeds with the CEPC on the announced timescale?*

There is a substantial overlap in physics goals and reach between CEPC and FCC-ee. From a scientific viewpoint, it would not make sense to build two large projects with similar goals and technical approaches. Furthermore, current geopolitical tensions may limit U.S. ability to participate in CEPC. Given the uncertainty in the execution of any plan and the scope of international participation, a CEPC inclusion in the next 5-year Plan of China should not immediately influence the ESG recommendations or CERN's direction to proceed with FCC-ee. ***The developments in China should be carefully monitored over the next several years and an appropriate strategy should be developed should China demonstrate its intent to move forward with CEPC construction.***

★ *if the US proceeds with a muon collider?*

The U.S. is not planning to build a muon collider[15] in the same time scale of FCC-ee construction. A demonstrator has been proposed for the 2030s, and the earliest likely launch of a U.S. based muon collider will likely follow the completion of the FCC-ee construction. P5 has suggested budget scenarios where both muon collider demonstrator and FCC-ee could be accommodated and minimal conflicts are expected, although, in some funding scenarios, the level of U.S. contribution to FCC-ee may be impacted. ***Given this, a planned muon collider effort in the U.S. should not influence the decision to move forward with the FCC-ee program at CERN.***

★ *If there are major new (unexpected) results from the HL-LHC or other HEP experiments?*

The physics realized at the HL-LHC and other HEP experiments will further deepen our understanding of the Standard Model and could well shape the direction of particle physics globally. However, waiting for HL-LHC physics to direct the future projects would lead to long gaps between HL-LHC and the next project and consequently result in an erosion of expertise. Furthermore, regardless of the HL-LHC results, it remains imperative to undertake a detailed study of the Higgs boson. ***Any new results from LHC and HL-LHC could help inform and fine-tune the relative operation at the different center of mass energies and optimize the transition to a higher energy collider.***

d) *Beyond the preferred option in 3a), what other accelerator R&D topics (e.g. high field magnets, RF technology, alternative accelerators/colliders) should be pursued in parallel?*

P5 has advocated for an off-shore Higgs Factory with a long-term vision for a 10 TeV pCM collider. If additional R&D is needed to develop a realizable near-term plan, that should be pursued with high priority. Beyond this, CERN, as well as the U.S., must play a leading role to ensure readiness for the realization of its long-term strategy, including investments in the following accelerator science and technology areas:

- High gradient R&D on superconducting radiofrequency (SRF) and normal-conducting radiofrequency (NCRF) technologies for a Muon Collider and $e^+e^-$ colliders.
- High-field magnet R&D, which is critical to realizing the FCC-hh program and a Muon Collider. This should include pursuing cost-effective implementation of high-field Nb3Sn and High Temperature Superconducting (HTS) technologies.
- Muon Collider design and technologies needed for an integrated design for a collider along with the design of a demonstrator facility needed to verify critical systems.
- R&D on advanced wakefield concepts are needed to develop an integrated design of a multi-TeV pCM collider.

e) *What is the prioritized list of alternative options if the preferred option set out in 3a) is not feasible (due to cost, timing, international developments, or for other reasons)?*

The P5 report supports a program that would explore Higgs physics on the timescale of the 2040's at CERN. If the FCC-ee is not feasible, an alternative consistent with the P5 vision is the ILC[14]. The ILC has been developed for several decades and has an advanced technical design. Both scientific and technical expertise also exists in the U.S. to advance an ILC or similar machine to its next phase.

Although not considered by P5, other alternatives, where U.S. scientific and technical expertise exist and benefit consideration, include LEP3[16], CLIC, and a lower energy hadron collider[17]. Noteworthy is the compelling physics case offered by a lower energy hadron collider, but which requires investments in developing cost-effective high field magnets.

*It is difficult to prioritize the alternatives at this time because none of them have completed Feasibility Studies[18] and the potential challenges facing the preferred option are currently unknown. The basis for any alternative plan should be a timely decision and realization of physics, and the smallest possible compromise in expected physics output compared to the preferred option. As with FCC-ee, U.S. participation will need to be consistent with the broader U.S. program.*

f) What are the most important elements in the response to e)? (The set of considerations in b should be used).

The response offered in (b) is applicable here as well.

**Technical capabilities and opportunities for engagement in Experiments**

U.S. groups at both universities and DOE national laboratories have extensive experience and capabilities in all areas of detector technologies. They have made leading contributions to the development, construction, and operation of many past and current HEP experiments at the Tevatron, PEP, CESR, B-Factories at SLAC and KEK, and the LHC, including the ongoing HL-LHC upgrades. The U.S. is well positioned to make leading contributions to the next generation collider experiments incorporating the latest state-of-the-art technological advancements and to drive the efforts in software and computing and other data analytic concepts such as artificial intelligence and machine learning (AI/ML). Furthermore, the U.S. EIC project has many similar detector requirements and may provide valuable experience with various technologies relevant to future Higgs Factories. Potential areas of U.S. contributions are listed below:

**Tracking Detectors:** The driving technical considerations for tracking and timing at the Higgs Factory include mass minimization, spatial precision, particle identification, and control of systematic uncertainties. These considerations match well with the experience, capabilities, and interests of the U.S. groups. Strong efforts exist already in the area of ASIC and solid state sensor design and fabrication, fast timing with LGADs, composite mechanics, carbon-fiber wire chambers, and gaseous tracking (particularly with straw tubes).

Natural areas for U.S. contributions to the development and construction of a future collider detector include the low mass solid state inner tracker, a gaseous outer tracker, either straws or drift chamber, and large area fast timing wrapper layers. U.S. teams are also capable of providing large-scale mechanical support structures made from carbon composites.

**Calorimeters**: New and existing technologies are being actively developed and optimized to support the unprecedented level of precision that can be achieved for future colliders. U.S. groups are positioned to make major contributions to the next generation of calorimeter detectors, including:

(i) *Dual Readout Calorimeters*, where recent work seeks to advance homogeneous crystal and fiber calorimetry with R&D paths into new materials, sensors, and readouts, supported by full detector simulations. U.S. efforts aim to simultaneously optimize EM and hadronic energy resolutions by including a transversely segmented homogeneous crystal EM layer with dual readout capabilities.
(ii) *Noble Liquid Calorimeters*, where U.S. institutes, with its experience in DZero, ATLAS and DUNE, are well placed to contribute to the cryogenic ASIC design and readout electronics. U.S. groups are pursuing unique turbine-like structures as a solution for forward calorimeters.
(iii) *Si-W calorimeters*, where U.S. groups co-led the design, construction, readout and reconstruction of the CMS forward high granularity calorimeter (HGCAL). U.S. efforts in the design of silicon-based calorimetry for ILC detectors, including a new ultra-high granularity concept using MAPs detectors to provide detailed imaging of EM showers pave the way for future developments.
(iv) *Tile Scintillators*: U.S. groups engaged in the development and production of the CMS HGCAL and ATLAS HCAL are positioned to engage in the design, fabrication, front end electronics and readout.

**Muon Spectrometer**: U.S.-based physicists have successfully designed, constructed, and operated large-scale muon detectors and have extensive expertise in building precision gaseous muon detectors. These include pressurized drift tubes, cathode strip chambers, and GEM technologies. The U.S. has also produced large detector systems at the Fermilab scintillator extrusion facility. Complementing these are state-of-the-art ultrafast front-end and back-end electronics, such as ASICs, high-density electronics boards, and FPGA-based readout and triggering systems. The U.S. is well positioned to design and construct high-precision, fast, robust, and cost-effective muon detector systems, drift tube chambers with fast scintillation strips with integrated SiPM readouts as well as MPGDs, which greatly enhance muon detection capabilities.

**AI and Microelectronics**: Among the most novel, interdisciplinary, and strategic areas of ongoing R&D efforts are the incorporation of novel AI/ML methods and advanced silicon microelectronics. These efforts are cross-cutting, with impact in each individual subsystem as well as on the overall detector design. Specific thrust areas where the U.S. can contribute include the development of ML-in-ASIC capabilities for ``smart detectors" capable of intelligent data handling and power management, the investigation of advanced topics such as analog compute and silicon photonics, the development of IP in modern technology nodes, and the use of AI-enhanced detector optimization techniques. Thorough investigation of these critical technologies allows for the development of sophisticated next-generation integrated detector concepts and the U.S. is well positioned to play a leading role in these developments.

**Trigger & DAQ**: The U.S. has significant expertise in the complete design of entire TDAQ systems stemming from their engagements in ATLAS, CMS and other experiments. The U.S. is well positioned to contribute to detector and data-flow simulations, algorithm and software development, board design, firmware development, verification, installation, commissioning, and operations. Ongoing R&D in exploring emerging technologies such as AI/ML, new memory and compute architectures, high-bandwidth data links, embedded and rad-hard FPGAs will allow for the U.S. to lead efforts in the next generation online data processing systems.

**Software & Computing:** As the largest provider of computing resources for the LHC experiments outside CERN, and due to its leading role in HEP software development, the U.S. is expected to have a significant role in software and computing for future colliders. The U.S. provides extensive computing resources for HEP data processing; for example, CMS and ATLAS have Tier-1 computing centers and several Tier-2 computing centers spread throughout the country. High-Performance Computing (HPC) facilities, including those at DOE and NSF supercomputing installations, provide additional resources for data processing and the development of large machine learning models. Computing sites receive support for grid infrastructure, software, and services from the Open Science Grid, and for network connectivity from ESnet. The U.S. institutes have also played a leading role in all aspects of software development at the LHC experiments, from reconstruction frameworks to distributed analysis platforms. The U.S. has extensive expertise and experience in core software development as well as physics software for simulation, reconstruction and end-user analysis, including event generation and detector modeling. Many of these are straightforward to translate for future Higgs Factories. The U.S. is therefore well positioned to engage in these efforts.

**Technical capabilities and opportunities for engagement in Accelerators**
Historically, the U.S. has been a leader in the development of both normal and superconducting linear collider designs as well as high luminosity colliding-beam storage rings. U.S. groups are in the process of constructing the Electron-Ion Collider (EIC), which will be one of the largest and most technically challenging colliding beam facilities in the world. The U.S. has technical accelerator expertise that could significantly contribute to any future Higgs Factory. A few examples of possible contributions are listed below for both FCC-ee and a possible Linear Collider (LC).

**Beam Dynamics:** There are many outstanding topics on FCC-ee or an LC that could be studied in collaboration with CERN and other partners. Accelerator scientists in the U.S. have developed codes for storage rings and linear colliders to model and simulate the beam-beam interaction, nonlinear dynamics, collective instabilities, spin dynamics, trajectory correction, and feedback systems. These codes are being used to design the EIC and were used to model the HL-LHC, the ILC, the KEK ATF2 Test Facility, and the CESR Damping Ring Test Facility. Many of these codes make use of massively parallel supercomputers or novel architectures (i.e., GPUs), or both. They can produce high-resolution, high-fidelity simulation results and would provide a complementary set of optimizations and verifications to existing tools at CERN for the future colliders.

Possible topics might include: a beam dynamics collaboration with CERN to adapt the optimizations used for the dynamic aperture in the EIC to study alternate paths to optimizing the FCC-ee, and to study combined effects from both beam-beam and machine impedance in the FCC-ee; a collaboration with partners to understand with simulations and experiments the present luminosity performance of SuperKEKB; a collaboration to model the dynamical behavior of the beams in FCC-ee or an LC where in FCC-ee the low revolution rate changes the expected feedback performance while the LC Final Focus Systems are very sensitive to accumulated aberrations, which were only partially understood at the ATF2 Test Facility; and development of AI/ML-based digital twin prototypes with applications to beam dynamics studies and commissioning.

**Diagnostics and Instrumentation:** Diagnostics and Instrumentation are critical for achieving rapid commissioning and reliable operation of the entire accelerator complex as well as for demonstrating the challenging accelerator parameters and conditions required for achieving ultimate luminosity performance and precision measurements of select systematic errors (e.g. on beam energy). The U.S. has expertise with most of the key technologies needed for either FCC-ee or an LC. However, in certain areas specialty diagnostics will need to be developed, for example for high-frequency measurements and control to facilitate top-up mode operations. This includes high-performance BPM systems, beam size diagnostics using synchrotron light or laser wires, beam loss monitors, diverse polarimeters, etc. Specialized diagnostics, required in the interaction regions, could align well with U.S. expertise.

**Magnets:** The U.S. has world-leading expertise in accelerator magnets, supporting critical technology advances and hardware production for HEP colliders. There are also important synergies between the Higgs Factory magnets and those of the EIC in the US. Specific expertise relevant to the FCC-ee and ILC Interaction Regions (IR) includes the development of Canted Cosine Theta and stress managed Cosine Theta magnets by the U.S. Magnet Development Program (MDP) collaboration, and the Direct Wind coil fabrication technique at BNL, which was successfully implemented in past projects and is expected to play a major role in the EIC. The U.S. is also well-positioned to contribute to the IR magnet integration and optimization, as was the case during the HL-LHC Accelerator Research Program and Design Study. Finally, there is significant U.S. expertise in both $Nb_3Sn$ that are applicable to special magnets and undulators, and in HTS technologies. Advances in HTS may lower the collider power requirements and possibly the costs.

**RF Systems:** The U.S. has technical expertise in SRF and normal conducting RF structures and systems. The U.S. was one of the major contributors to the ILC 1.3 GHz SRF development and is presently collaborating with CERN on the 800 MHz SRF cavities and cryomodules needed for the FCC-ee. Cavity processing techniques with which the U.S. has extensive and pioneering experience, such as impurity-doping and low- to mid-temperature baking, as well as alternative superconductors (e.g., $Nb_3Sn$), are necessary to meet the ambitious SRF cavity performance goals of the FCC-ee. Similarly, distributed coupling and possible LN-cryogenic operation of normal conducting structures could have application to either an LC or the FCC-ee injector.

The U.S. also has expertise with both vacuum tube RF power sources and solid-state RF power sources. Vacuum tube RF sources are needed for FCC-ee and an LC. The tube industry has been in decline and developing international vendors will be critical. Increasing the efficiency of the RF sources and improving manufacturing techniques would also have a big impact on the FCC-ee and LC. Similarly, increasing the efficiency of solid-state amplifiers using LDMOS or GaN has the potential for a significant impact on the sustainability of future accelerators.

**MDI:** The Machine-Detector Interface (MDI) is a critical part of any colliding beam accelerator. Successful development of the MDI requires integration of the MDI components including magnets, diagnostics, collimation and masks, vacuum chamber, and component supports. It requires balancing the accelerator beam physics, experimental detector requirements, and detailed engineering of the system. U.S. institutes manage some of the highest-performing Particle-in-Cell (PIC) codes in the world and intend to contribute novel tools and workflows for modeling beam-beam interactions and detector backgrounds with PIC codes for fast, high-resolution modeling of the Interaction Region. The U.S. also has experience developing the MDI for past colliders. It was a leader in the development of the ILC MDI and is in the process of designing the MDI for the EIC. The U.S. would be in a strong position to collaborate on specific subsystems or the overall integration of the MDI systems for one or more of the IRs. Such an effort would logically tie with US institutional experience in the design and operation of one or more of the experimental detectors.

**Engineering and Infrastructure:** The U.S. has engineering resources and design experience that can be applied to either FCC-ee or an LC. The U.S. is already collaborating on the design of the FCC-ee surface buildings and could extend that effort within both surface buildings and underground work. The U.S. can also provide access to existing test facilities and other necessary space for R&D as well as data collection and processing. The U.S. could also manage industrial partnerships for component manufacturing and technology transfer with an emphasis on efficiency, sustainability, and large-quantity production. The U.S. also has engineering experience working on critical subsystems such as the MDI which can be applied to either the FCC-ee or an LC.

**Sustainability and Integration:** Improving sustainability is critical for all future accelerators as it can reduce the overall cost of a project and develop technologies and techniques that have broader impacts to other future scientific endeavors. There are many areas where the U.S. could contribute ranging from the development of high-efficiency RF power sources and use of HTS technology to engineering the site power distribution and cooling systems. The U.S. could also contribute to the activities of integration and coordination of the subtopics, and optimization of the design.

## Perspectives of Early Career Researchers

Through two open sessions[6] at SLAC and Fermilab and through an online survey directed at members of the Higgs Factory community, the early career researchers (ECR) community contributed input and perspectives on the European Strategy Group (ESG) questions. In general, the ECR perspective is well-aligned with the positions highlighted in this document.

The foremost priority of Higgs Factory ECRs is the prompt establishment of a clear strategy and decision for CERN's future. Among early career scientists in the US, the *physics potential* of a future collider and the need for a timely decision was the key motivation for selecting FCC-ee in response to ECFA's item 3a)[1]. The ECR community also highlighted career stability, research funding, and long-term prospects for the high-energy physics field as influential factors in supporting the FCC.

Beyond the next collider, the realization of a 10 TeV pCM energy collider on the shortest possible time scale is a high priority for the ECR community, and therefore a robust R&D effort should be supported targeting this goal. This was another important factor driving ECR support for FCC-ee. While some questions were raised about coordinating resources and timelines between the Higgs factory, different 10 TeV pCM energy collider designs, and their eventual implementation, the ECR community strongly endorses the FCC program, beginning with FCC-ee, as CERN's next major endeavor.

## Summary

The U.S. is enthusiastic for a Higgs Factory as the next major collider and strongly supports FCC-ee, intending to collaborate on its construction and physics exploitation if it is chosen as the next major research infrastructure project at CERN. The U.S. recommends that appropriate steps be taken to develop a realizable plan in the coming 3-5 years. The U.S. would also support an LC if the CERN Council approves such a project in a timely manner. The U.S. eagerly awaits a CERN Council decision and looks forward to partnering with CERN on the next future collider project.